\documentstyle[12pt,epsfig]{article}             
\newcommand{\beq}{\begin{equation}}             
\newcommand{\eeq}{\end{equation}}               
\newcommand{\bqry}{\begin{eqnarray}}            
\newcommand{\eqry}{\end{eqnarray}}              
\newcommand{\bqryn}{\begin{eqnarray*}}          
\newcommand{\eqryn}{\end{eqnarray*}}            
\newcommand{\preprint}[1]{\begin{table}[t]      
            \begin{flushright}                  
            \begin{large}{#1}\end{large}        
            \end{flushright}                    
            \end{table}}                        
\newcommand{\PD}[2]                             
    {\frac{\partial^{#2}}{\partial #1^{#2}}}    
\begin{document} 
\preprint{LA-UR-00-633} 
\title{Dislocation-Mediated Melting: \\ 
The One-Component Plasma Limit} 
\author{\\ Leonid Burakovsky\thanks{E-mail: BURAKOV@LANL.GOV} \
and Dean L. Preston\thanks{E-mail: DEAN@LANL.GOV}
 \\  \\ 
Los Alamos National Laboratory \\ Los Alamos, NM 87545, USA }
\date{ }
\maketitle
\begin{abstract} 
The melting parameter $\Gamma _m$ of a classical one-component plasma is 
estimated using a relation between melting temperature, density, shear 
modulus, and crystal coordination number that follows from our model of 
dislocation-mediated melting. We obtain $\Gamma _m=172\pm 35,$ in good 
agreement with the results of numerous Monte-Carlo calculations. 
\end{abstract}
\bigskip 
\centerline{{\it Key words:} melting, dislocation, one-component plasma, 
Monte-Carlo simulations} 

\bigskip

\section{Introduction} 

The classical one-component plasma (OCP) is an idealized system of mobile 
ions of charge $Ze,$ number density $n,$ and temperature $T,$ immersed in 
a neutralizing background of uniform charge density $-Zne.$  
The OCP is realized in nature only at the enormous densities occuring in 
white dwarfs and neutron stars. The thermodynamics of the classical OCP is 
completely described in terms of the dimensionless coupling parameter 
\cite{reviews} 
\beq
\Gamma =\frac{(Ze)^2}{ak_BT}, 
\eeq 
where $a=(3/4\pi n)^{1/3}$ is the Wigner-Seitz radius. In the quantum regime, 
one more parameter, $a$ or $T,$ is needed to characterize the system.
Melting of a classical OCP occurs at a fixed value, $\Gamma _m,$ of the 
plasma coupling parameter. When $\Gamma >\Gamma _m,$ a OCP is either a glass 
\cite{glass}, or it has a bcc crystal structure provided that it is subject 
to only hydrostatic stress. The evaluation of $\Gamma _m$ for melting from the 
bcc structure has been the subject of extensive Monte Carlo (MC) calculations 
[3-16] employing the Ewald potential, which yields data pertinent to an 
infinite system from simulations using only a finite number of particles 
confined to a cubic computational cell with periodic boundary conditions. 
By fitting simple functional forms, guided by theory, to 
the measured excess potential energy per particle for both liquid and solid 
phases of the OCP, it is possible to obtain the Helmholtz free energy as a 
function of $\Gamma .$ The intersection of the liquid and solid free-energy 
curves gives the value of the melting parameter $\Gamma _m.$ In their 
pioneering study \cite{BST}, Brush, Sahlin and Teller observed melting in a 
32-particle system at $\Gamma _m\approx 125.$ 
Subsequently Hansen \cite{Hansen} and Pollock and Hansen \cite{PH} 
followed with an improved calculation and found $\Gamma _m=155\pm 10.$ 
Van Horn \cite{VH} used the empirical Lindemann melting criterion to obtain 
$\Gamma _m=170\pm 10.$ Other MC studies resulted in the following values of 
$\Gamma _m:$ 144 \cite{DW}, $168\pm 4$ 
\cite{DWR}, $178\pm 1$ \cite{SDDW}, 
$180\pm 1$ \cite{OI}, 178 
\cite{SDWS}, 172 \cite{Dubin}, and 173 \cite{FH1}. Values of very similar 
magnitude have been obtained in MC simulations of a strongly-coupled 
screened-Coulomb (Yukawa) system in the limit of zero screening: 171 
\cite{FH2} and 171.8 \cite{FHD}. Recent path-integral MC simulations of the 
OCP \cite{JC} give $\Gamma _m=175.$ Hence, numerous MC studies suggest that 
$\Gamma _m\simeq 170-180$ for the classical bcc OCP.

In this paper we calculate $\Gamma _m$ using a melting relation obtained 
from our model of dislocation-mediated melting \cite{prev1,prev2}. Before 
proceeding with the calculation of $\Gamma _m,$ we briefly recapitulate the 
main ideas and assumptions of our melting model. As first proposed 
by Mott \cite{Mott}, dislocations are assumed to be the basic degrees of 
freedom underlying the melting transition. Dislocation interactions beyond 
a distance of order the mean dislocation separation are assumed negligible 
because of screening, and steric interactions are ignored. Accordingly, 
dislocations are taken to be non-interacting and therefore uncorrelated, 
and are modeled as lines lying along the nearest-neighbor links of the 
lattice. The links coincide with the shortest perfect-dislocation Burgers 
vectors, which have magnitude $b.$ The dislocation configurations (Brownian, 
self-avoiding, open, closed, etc.) are parametrized by a single parameter 
$q>1,$ in terms of which the mean dislocation length is given by $\langle 
L\rangle =4qb/(q-1).$ In addition to $q,$ the partition function depends on 
the temperature-dependent effective dislocation line tension, that is, the 
energy cost to create unit length of dislocation at temperature $T.$ The 
effective line tension vanishes at the critical temperature $k_BT_{cr}=
\sigma b/\ln (z-1).$ Here $z$ is the coordination number of the lattice 
and $\sigma ,$ which we discuss in more detail below, is the $\rho $-dependent 
self-energy per unit length, $\rho $ being the dislocation density. 
Dislocations proliferate as $T_{cr}$ is approached from below, while 
at temperatures just above $T_{cr}$ the partition function diverges, an 
indication that a new phase appears. So $T_{cr}$ corresponds to a phase 
transition, namely melting, and we identify $T_{cr}$ with the melting 
temperature, $T_m.$ A full defect theory of melting, a version of which 
is presently available \cite{KleinII}, would have to include the effects 
of both dislocations and disclinations. In our model we ignore the effects 
of disclinations under the assumption that they will produce only small 
changes of the order of 10\% to the melting temperature. 

Under the assumption that dislocation strain fields are screened away at 
distances beyond the mean interdislocation spacing, the self-energy per 
unit length is given by \cite{HL} 
\beq 
\sigma =\frac{1-\nu /2}{1-\nu }\;\frac{Gb^2}{4\pi }\;\!\ln \left( \frac{
\alpha R}{b}\right) =\frac{1-\nu /2}{1-\nu }\;\frac{Gb^2}{8\pi }\;\!\ln 
\left( \frac{\alpha ^2}{4b^2\rho }\right) , 
\eeq 
where $2R\approx 1/\sqrt{\rho }$ is the mean distance between dislocations, 
$\nu $ is the Poisson ratio, $G$ is the shear modulus and $\alpha $ accounts 
for non-linear effects in the dislocation core. Many authors \cite{many} have 
successfully used this $\ln (1/\rho )$ form for $\sigma ,$ so we chose it 
as well, even though it has not been thoroughly investigated theoretically. 
Careful derivations \cite{derivations} have been carried out only for nearly 
parallel dislocations. However, the $\ln (1/\rho )$ form is expected to hold 
in a three-dimensional ensemble of non-directed dislocations provided the 
mean dislocation length is much larger than the mean distance between 
dislocations, that is, $\langle L\rangle \sqrt{\rho }\gg 1.$ In our model 
the $\ln (1/\rho )$ self-energy leads to a dislocation free energy $F=-a_1
\rho \ln \rho -a_2\rho -a_3\rho ^{a_4},$ and the $\rho \ln \rho $ term 
results in a first-order melting transition. 

We obtain the following melting relation: 
\beq 
k_BT_m=\frac{1-\nu /2}{1-\nu }\;\frac{\lambda G(T_m)v_{WS}(T_m)}{8\pi 
\ln (z-1)}\;\!\ln \left( \frac{\alpha ^2}{4b^2\rho (T_m)}\right) . 
\eeq 
Here $v_{WS}$ is the Wigner-Seitz volume, $\lambda \equiv b^3/v_{WS}$ is 
a geometric factor characterizing the lattice, and $\rho (T_m)$ is the 
dislocation density at melt. Note that the factor $\ln (z-1)$ explicitly 
accounts for the influence of crystal structure on melting. This melting 
relation plus experimental data on over half the elements in the periodic 
table gives $b^2\rho (T_m)=0.61\pm 0.20$ \cite{prev2}. 

In ref.\ \cite{prev1} we applied Eq.\ (3) to the zero-pressure elemental 
data for more than half of the periodic table and found that it is accurate 
to 17\%. Here we investigate the validity of this relation in the OCP limit 
by using it to calculate the value of $\Gamma _m$, which is then compared 
to the available MC data. 

Calculation of $\Gamma _m$ from Eq.\ (3) requires that we make the reasonable 
assumption that $\alpha ^2/b^2\rho (T_m)$ is a pressure-independent constant. 
Then we can estimate this quantity for the OCP from zero-pressure data on the 
alkali metals. It is well known that the deviations of alkali-metal Fermi 
surfaces from perfect spheres are of order 1\% or less, clear evidence that 
the valence electrons are very nearly free. In addition, the ratio of ionic 
radius to half the interatomic distance increases from 0.4 in Li to only 0.7 
in Cs \cite{Fuchs}, hence the overlap between alkali ions is small, and so 
to a good approximation the ions are effectively point charges. With respect 
to many of its physical properties (third-order elastic constants are one 
exception \cite{SGT}), an alkali metal can be regarded as a bcc lattice 
of point positive ions in a uniform background of free electrons, i.e., 
a one-component plasma. 

\section{Analysis of alkali metal data} 

Let us first discuss the temperature dependencies of $G$ and $v_{WS},$ 
since their values to be used in Eq.\ (3) should be those at $T=T_m,$ 
not the measured values at room temperature.

The fixed-pressure ratio of Wigner-Seitz volumes at $T_m$ and $T=0$ is equal 
to $1+\beta T_m,$ where $\beta $ is the volume expansivity. At $p=0,$ $\beta $
is typically of order $10^{-5}$ K$^{-1},$ and melting temperatures are at 
most about 4000 K, so $v_{WS}$ changes by only a few percent between $T=0$ 
and $T_m.$ We can therefore always use room-temperature values for $v_{WS}.$ 

In contrast to $v_{WS},$ the dependence of $G$ on $T$ is not necessarily weak. 
Its $T$-dependence involves two characteristic temperatures, namely the Debye 
temperature, $T_D,$ and the melting temperature. $G$ is always monotonically 
decreasing with $T,$ and is nonlinear for $T\stackrel{<}{\sim }T_D$ and 
linear from $T_D$ to $T_m.$ An accurate representation of $G(T)$ at fixed 
density is achieved by ignoring the low-temperature non-linearity and 
approximating $G(T)$ as a linear function of the reduced temperature $T/T_m$ 
with the correct value $G(0)$ at $T=0$ \cite{PW}: 
\beq 
G(T)=G(0)\left( 1-\gamma \frac{T}{T_m}\right) . 
\eeq 
This straight-line representation turns out to be quite accurate: 
the maximum deviation of the data from the corresponding fitted lines 
is $\sim 11$\% for the 22 metals analyzed in \cite{PW}. 

Hence, as follows from Eq.\ (3), 
\beq 
\frac{1}{\lambda \ln (\alpha ^2/4b^2\rho (T_m))}=\frac{1-\nu /2}{1-\nu }
\;\frac{1-\gamma }{2}\;\frac{G(0)v_{WS}}{4\pi k_BT_m\ln 7}, 
\eeq 
where we have taken $z=8.$ For our analysis, we use Li, Na, K, Rb, and Cs, 
and omit Fr for which lattice constant data are not available. For the 
remaining 5 alkali metals, $v_{WS}=a^3/2,$ where the values of the lattice 
constant, $a$, are taken from \cite{LB-a}. The values of both $G(0)$ and 
$T_m$ come from \cite{Gschn}. We take the values of $\gamma $ from 
ref.\ \cite{PW} for Na, K and Rb, and for Li and Cs we use $\gamma =0.23,$ 
which is the average value over the 22 metals analyzed in \cite{PW}. The 
corresponding values of $\nu $ are taken from ref.\ \cite{Gschn}. Averaging 
over the five alkali metals gives 
\beq 
\frac{1}{\lambda \ln (\alpha ^2/4b^2\rho (T_m))}=0.385\pm 0.052, 
\eeq 
where the error is the root-mean-square deviation. 

%

\section{OCP melting parameter $\Gamma _m$} 

We first consider the Poisson ratio in the OCP limit. In terms of $G$ and 
the bulk modulus, $B,$ the Poisson ratio is given by \cite{Gschn} 
\beq 
\nu =\frac{3B-2G}{2(3B+G)}. 
\eeq 
We approximate $B^{{\rm OCP}}$ by the bulk modulus of the electron gas since 
the negative electrostatic (Madelung) contribution never exceeds 10\% of the 
bulk modulus of the gas. The variation of $B^{{\rm OCP}}$ with density changes 
from a $n^{5/3}$ dependence in the non-relativistic case to a $n^{4/3}$ 
dependence in the extreme relativistic limit, whereas the OCP shear modulus 
always varies as $n^{4/3}.$ Hence, in a non-relativistic gas, $B/G\sim n^{1/3}
\gg 1,$ so $\nu \rightarrow 1/2,$ as seen from Eq.\ (7). Although $B$ and $G$ 
both vary as $n^{4/3}$ in the extreme relativistic limit, we find $B/G\approx 
1000\;Z^{-2/3},$ so again $B/G\gg 1$ and $\nu \approx 1/2.$ 

Hence, as follows from Eqs.\ (1),(3) and (6) with $\nu =1/2,$ 
the OCP melting parameter is given by 
\beq 
\Gamma _m=(0.385\pm 0.052)\;\!8\pi \ln 7\;\frac{2}{3}\;\frac{
(4\pi /3)^{1/3}(Ze)^2n^{4/3}}{G^{{\rm OCP}}(\Gamma _m)}, 
\eeq 
where we have used $v_{WS}=1/n.$ 

The bcc OCP elastic constants were recently obtained by Ogata and Ichimaru, 
using MC simulations \cite{OI2}, as functions of $\Gamma .$ However, the 
formula for the effective shear modulus used in ref.\ \cite{OI2}, 
\beq 
G_{{\rm eff}}=\frac{c_{11}-c_{12}+3c_{44}}{5}, 
\eeq 
is in fact the Voigt (upper) bound \cite{V} on the shear modulus, and 
therefore does not give the correct value of $G,$ which is known to always 
lie between the Voigt and the Reuss (lower) \cite{R} bounds. 

An analysis by Kr\"{o}ner \cite{Kron} shows that successively narrower bounds 
can be placed on the shear modulus as the degree of disorder in grain 
orientation increases. In the limit of perfect disorder, the shear modulus 
can be obtained as a root of a cubic equation with coefficients that depend 
on the single-crystal elastic constants. In the case of the OCP, where 
the shear modulus is down by a factor of $n^{1/3}$ from the bulk modulus, 
the cubic equation reduces to a quadratic with only one positive real root: 
\beq 
G=\frac{1}{6}\left[ c_{44}+\sqrt{ c_{44}^2+12\;\!(c_{11}-c_{12})
\;\!c_{44}}\right] . 
\eeq 
In Table I we present the values of the elastic constants from ref.\ 
\cite{OI2} and the correct values of $G^{{\rm OCP}}$ as calculated from 
Eq.\ (10). 
 \\ 

\begin{center}
\begin{tabular}{|c|l|l|l|}
\hline 
$\Gamma $ & $(c_{11}-c_{12})/2$ & $c_{44}$ & $G^{{\rm OCP}}$   \\
\hline 
$\infty $ & 0.02454  & 0.1827   & 0.0930      \\ 
    800   & 0.024(2) & 0.174(1) & 0.089(12)   \\ 
    400   & 0.025(2) & 0.167(1) & 0.087(11)   \\ 
    300   & 0.025(3) & 0.157(4) & 0.084(19)   \\ 
    200   & 0.019(3) & 0.12(1)  & 0.064(28)   \\ 
\hline 
\end{tabular}
\end{center}
\centerline{Table I. The elastic constants and shear modulus,}
\centerline{in units of $(4\pi /3)^{1/3}(Ze)^2n^{4/3}.$} 
\vspace*{0.5cm} 

Let us again assume a linear temperature dependence of $G^{{\rm OCP}}$ 
on $T\propto 1/\Gamma :$ 
\beq 
G^{{\rm OCP}}(\Gamma )=
G^{{\rm OCP}}(\infty )\left( 1-\frac{\eta }{\Gamma }\right) . 
\eeq  
Fitting the values in Table I to this linear formula, and taking into account 
their uncertainties, we obtain \cite{MM} 
\beq 
\eta =36.7\pm 30.4.
\eeq 

Finally, we evaluate the OCP melting parameter $\Gamma _m$ in the framework 
of melting as a dislocation-mediated phase transition. As follows from 
Eqs.\ (8), and (11),(12) with the value of $G^{{\rm OCP}}(\infty )$ from 
Table I, 
\beq 
\Gamma _m=
172\pm 35. 
\eeq 
This value is in good agreement with the available data from MC simulations, 
albeit with 20\% uncertainty. We note that most $(\approx 2/3)$ of 
this uncertainty comes from the uncertainty in the value of $\eta .$ 

The OCP value of the parameter $\gamma $ defined in Eq.\ (4) is simply related 
to $\eta $ and $\Gamma _m:$ 
\beq 
\eta \equiv \gamma ^{{\rm OCP}}\Gamma _m. 
\eeq 
From Eqs.\ (12) and (13) we get $\gamma ^{{\rm OCP}}=
0.21\pm 0.18,$ which is consistent with the value of $\gamma $ 
at $p\sim 0,$ namely 0.23 \cite{PW}.

\section{Concluding remarks} 

Our central value for $\Gamma _m,$ that is 172, agrees well with the more 
recent MC results. Two-thirds of the 20\% uncertainty in this value is 
attributable to the error in the MC-calculated temperature dependence of 
the OCP single-crystal elastic constants. 

Our previous study of the melting curves of 18 elements \cite{prev3} revealed 
that the melting relation (3) is in good agreement with data up to pressures 
$\sim 100-200$ GPa. Here we have demonstrated that Eq.\ (3) also holds in a 
classical OCP. These successful comparisons of Eq.\ (3) with experimental data 
and MC calculations suggest, but of course do not by themselves prove, that 
melting is a dislocation-mediated phase transition.  

\section*{Acknowledgements} 

We wish to thank T.\ Goldman and R.R.\ Silbar for very valuable discussions 
during the preparation of this work, and M.M.\ Brisudov\'{a} for her help 
with a least-squares fitting.


\bigskip
\bigskip
\begin{thebibliography}{9} 
\bibitem{reviews} M. Baus and J.P. Hansen, Phys. Rep. {\bf 59} (1980) 1 \\ 
S. Ichimaru, Rev. Mod. Phys. {\bf 54} (1982) 1017 \\ 
S. Ichimaru, H. Iyetomi and S. Tanaka, Phys. Rep. {\bf 149} (1987) 91 
\bibitem{glass} S. Ichimaru, H. Iyetomi, S. Mitake and N. Itoh, Ap. J. 
{\bf 265} (1983) L83 
\bibitem{BST} S.G. Brush, H.L. Sahlin and E. Teller, J. Chem. Phys. {\bf 45} 
(1966) 2102 
\bibitem{Hansen} J.P. Hansen, Phys. Rev. A {\bf 8} (1973) 3096 
\bibitem{PH} E.L. Pollock and J.P. Hansen, Phys. Rev. A {\bf 8} (1973) 3110 
\bibitem{VH} H.M. Van Horn, Phys. Lett. A {\bf 28} (1969) 706 
\bibitem{DW} H.E. DeWitt, Phys. Rev. A {\bf 14} (1976) 1290 
\bibitem{DWR} H.E. DeWitt and Y. Rosenfeld, Phys. Lett. A {\bf 75} (1979) 79 
\bibitem{SDDW} W.L. Slattery, G.D. Doolen and H.E. DeWitt, Phys. Rev. A 
{\bf 21} (1980) 2087, {\bf 26} (1982) 2255 
\bibitem{OI} S. Ogata and S. Ichimaru, Phys. Rev. A {\bf 36} (1987) 5451 
\bibitem{SDWS} G.S. Stringfellow, H.E. DeWitt and W.L. Slattery, Phys. Rev. A 
{\bf 41} (1990) 1105 
\bibitem{Dubin} D.H.E. Dubin, Phys. Rev. A {\bf 42} (1990) 4972 
\bibitem{FH1} R.T. Farouki and S. Hamaguchi, Phys. Rev. E {\bf 47} (1993) 4330 
\bibitem{FH2} R.T. Farouki and S. Hamaguchi, J. Chem. Phys. {\bf 101} (1994) 
9885 
\bibitem{FHD} S. Hamaguchi, R.T. Farouki and D.H.E. Dubin, Phys. Rev. E 
{\bf 56} (1997) 4671 
\bibitem{JC} M.D. Jones and D.M. Ceperley, Phys. Rev. Lett. {\bf 76} (1996) 
4572 
\bibitem{prev1} L. Burakovsky and D.L. Preston, Solid State Comm. {\bf 115} 
(2000) 341 
\bibitem{prev2} L. Burakovsky, D.L. Preston and R.R. Silbar, Phys. Rev. B 
{\bf 61} (2000) 15011 
\bibitem{Mott} C. Mott, Proc. R. Soc. London, Ser. A {\bf 215} (1952) 1
\bibitem{KleinII} H. Kleinert, {\it Gauge Fields in Condensed Matter,} 
(World Scientific, Singapore, 1989), Vol.\ II 
\bibitem{HL} J.P. Hirth and J. Lothe, {\it Theory of Dislocations,} 2nd ed., 
(Krieger, Malabar, FL, 1992) 
\bibitem{many} S. Mizushima, J. Phys. Soc. Japan {\bf 15} (1960) 70 \\ 
R.M.J. Cotterill, J. Cryst. Growth {\bf 48} (1980) 582 \\ 
J. Kierfeld and V. Vinokur, Dislocations and the critical endpoint of the 
melting line of vortex line lattices, cond-mat/9909190 
\bibitem{derivations} T. Yamamoto and T. Izuyama, J. Phys. Soc. Japan {\bf 57} 
(1988) 3742 \\ G.F. Sarafanov, Phys. Solid State {\bf 39} (1997) 1403 
\bibitem{Fuchs} K. Fuchs, Proc. Roy. Soc. London A {\bf 153} (1936) 622 
\bibitem{SGT} T. Suzuki, A.V. Granato and J.F. Thomas, Jr., Phys. Rev. 
{\bf 175} (1968) 766 
\bibitem{PW} D.L. Preston and D.C. Wallace, Solid State Comm. {\bf 81} (1992) 
277 
\bibitem{LB-a}  Landolt-B\"{o}rnstein Numerical Data and Functional 
Relationships in Science and Technology, New Series, Eds. K.-H. Hellwege and 
O. Madelung, Group III: Crystal and Solid State Physics, (Springer-Verlag, 
Berlin), Vol. 7 (1971), Vol. 14 (1988), Vol. 25 (1991) 
\bibitem{Gschn} K.A. Gschneidner, Jr., in {\it Solid State Physics, Advances 
in Research and Applications,} Eds. F. Seitz and D. Turnbull, (Academic Press,
New York, 1965), Vol.\ 16, p.\ 275 
\bibitem{OI2} S. Ogata and S. Ichimaru, Phys. Rev. A {\bf 42} (1990) 4867 
\bibitem{V} W. Voigt, {\it Lehrbuch der Kristallphysik,} (Teubner, Leipzig, 
1928), p.\ 962 
\bibitem{R} A. Reuss, Z. Angew. Math. Mech. {\bf 9} (1929) 55 
\bibitem{Kron} E. Kr\"{o}ner, Z. Phys. {\bf 151} (1958) 504, Eng. Mech. Div. 
Amer. Soc. Civ. Eng. {\bf 108} (1980) 899; 
B.K.D. Gairola and E. Kr\"{o}ner, Int. J. Eng. Sci. {\bf 19} (1981) 865 
\bibitem{MM} M.M. Brisudov\'{a}, private communication 
\bibitem{prev3} L. Burakovsky, D.L. Preston and R.R. Silbar, Analysis 
of dislocation mechanism for melting of elements: Pressure dependence, 
cond-mat/0005118, J. Appl. Phys., {\it in press} 
\end{thebibliography}
\end{document}